\documentclass[runningheads]{llncs}
\usepackage{graphicx}
\usepackage{enumitem}
\usepackage{ulem}
\usepackage{algorithm}
\usepackage{algorithmic}

\begin{document}
\title{WebMap – Large Language Model-assisted Semantic Link Induction in the Web}
\titlerunning{WebMap – LLM-assisted Semantic Link Induction in the Web}
%
\author{Shiraj Pokharel\inst{1}\orcidID{0009-0003-5777-2749} \and
Georg P. Roßrucker\inst{2}\orcidID{0009-0007-2430-6687} \and
Mario M. Kubek\inst{1}\orcidID{0000-0003-2641-2065}}
%
\authorrunning{S. Pokharel et al.}
%
\institute{Georgia State University, Atlanta, GA, USA\\
\email{\{spokharel3,mkubek\}@gsu.edu}\and
FernUniversität in Hagen, Hagen, Germany\\
\email{g.rossrucker@gmail.com}}
%
\maketitle              
\begin{abstract}
Carrying out research tasks is only inadequately supported, if not hindered, by current web search engines. This paper therefore proposes functional extensions of \textit{WebMap}, a semantically induced overlay linking structure on the web to inherently facilitate research activities. These add-ons support the dynamic determination and regrouping of document clusters, the creation of a semantic signpost in the web, and the interactive tracing of topics back to their origins.

\end{abstract}
\section{Initial Situation}

Web search engines exploit the explicit linking structure of the World Wide Web (WWW) to determine the relationships between web documents and assess the relevance and authority of content. Typically, hyperlinks are intentionally created and strategically placed by human efforts. However, it would be beneficial to also use semantically induced links between web documents and their content fragments to indicate topical relations and topically group potentially relevant web search results. This approach could facilitate labor-intensive research processes by automating the identification of relevant connections and topics.

Furthermore, as web search engines do not topically group potentially relevant web search results, conducting research using them constitutes a labor-intensive task as it involves having to manually inspect and evaluate the returned results. Therefore, it might be beneficial to return clusters of conceptually similar and related documents that match a query's intent, too. 

To this end, in \cite{Rossrucker:24}, we introduced the \textit{WebMap}, a novel solution to extending the existing linking structure of a hyperlinked network of text documents such as the WWW by a peer-to-peer-based semantic overlay, which induces and represents a distributed graph structure. 

The paper's remainder introduces conceptual extensions to \textit{WebMap}, aimed at enhancing support for research tasks. These extensions, though still in development, are providing functionalities such as dynamic determination and regrouping of document clusters, creation of semantic signposts, and interactive tracing of topics. By conceptualizing these advancements, the paper lays the groundwork for future development and implementation within \textit{WebMap} to better facilitate research activities on the Web.

\section{WebMap's Architecture}
The main idea of \textit{WebMap} is to embed a semantic and meaningful linking mechanism into the existing Web to make navigation and search independent of the existing -- rather chronologically evolved -- link structure. This is crucial because classical hyperlinks typically point to existing (older) content, putting new content at a disadvantage in terms of discoverability.

To achieve this, the global overlay linking structure is designed as a network of so called \textit{Cluster Files}, generated and provided by the participating peers (web servers). Cluster files are identified by meaningful terms (text-representing centroids, TRCs \cite{Kubek:16}) and consist of two sets of hyperlinks:

\begin{itemize}
  \item a set of links referring to documents (leaves) that are related within the given cluster’s \textit{context}, and 
  \item a set of links pointing to semantically related clusters, realizing traversable, bi-directional edges.
\end{itemize}

\begin{figure}
    \centering
    \includegraphics[width=0.55\linewidth]{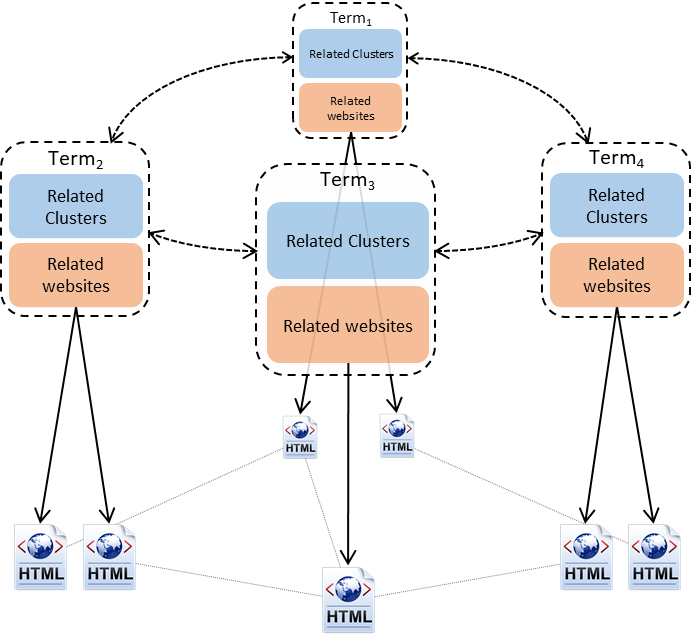}
    \caption{WebMap's architecture}
    \label{fig:WebMapStructure}
\end{figure}

Fig.~\ref{fig:WebMapStructure} illustrates how the overlaying linking structure of the \textit{WebMap} extends the present linking structure of the underlying Web.

\section{Improving WebMap}
\subsection{Language Modeling and Large Language Models (LLMs)}
Let's take a representative sentence - ``A German Shepherd is a [blank]". Most likely, our guess for [blank] is ``dog", which means that the context provided does not refer to a "human shepherd with German nationality". Similarly, assigning a guess or probability to every possible word via mathematical reasoning and computation is called language modeling.
Further, language models developed via deep neural networks and the Transformer \cite{Vaswani:17} architecture with possibly hundreds of billions of programmable connections (called ``parameters") are referred to as Large Language Models (LLMs). Below, we briefly touch upon two of the most successful LLM-families recently.

\subsubsection{Generative Pre-trained Transformers (GPT)} 
Released by OpenAI, GPTs  are types of decoder-only Transformer \cite{Vaswani:17}-based language models. GPT-1 and GPT-2 are open source while GPT-3 \cite{Brown:20} and GPT-4 \cite{GPT4:23} are closed source models \cite{Minaee:24}. GPT-3 is the model which excels in tasks that can be provided in natural human language like translation.
GPT-4 is a multi-model which can produce text-based outputs from image or text-based inputs.

\subsubsection{LLaMA}
Released by Meta, LLaMA \cite{laama1:23} uses the transformer architecture of
GPT-3 \cite{Brown:20}, with some modifications in the activation function, positional embeddings and layer-normalization. The latest release is LLaMA-2 \cite{laama2:23}. Thus far, LLaMA models are open-sourced by Meta where the model weights are made publicly available for non-commercial study and research purposes.

\subsection{Neural Induction of Local Term Proximity Graphs}

To derive the globally valid cluster assignment for documents and the necessary relations between clusters, peers of the \textit{WebMap} make use of individual co-occurrence graphs \cite{Jin:07} that are induced by local text documents and capture simple syntagmatic term relations \cite{Biemann:22}. Extracting \textit{significant} term relations and \textit{meaningful} topical connections between documents and cluster files therefore greatly depends on the available number, quality, and context of the individual peers' documents.

To obtain more meaningful document assignments and cluster associations in a harmonized manner that is commonly accepted and verifiable, we propose to change the underlying mechanism from using co-occurrence graphs to local term proximity graphs induced by LLMs such as BERT \cite{Devlin:19} and its variants, GPT-3 \cite{Brown:20}, GPT-4 \cite{GPT4:23}, LLaMA-1 \cite{laama1:23}, LLaMA-2 \cite{laama2:23}, Mistral 7B \cite{Jiang:23}, etc. Following the principle of distributional semantics \cite{Biemann:22}, LLMs are able to convert input words into (sub)tokens and then into contextualized vector embeddings (high-dimensional real-valued vectors) that capture the words' meanings in context.

To induce the required local term proximity graphs, we hence propose to generally compare the word embeddings of selected and meaningful textual elements such as nouns and proper nouns that appeared together in the local documents using the cosine similarity measure. We will refer to them as terms in the following considerations. The following two approaches to determining the degree of term similarity make sense here: 

\begin{itemize}
  \item In the first approach, we compute the embeddings for each occurrence of a selected term in a document. Then, we average these embeddings to create a single term representation for the entire document. After that, we compare the averaged embeddings of each pair of terms using the cosine similarity measure to find out how similar they are in meaning. This method helps us understand their overall semantic relationships within the document.
  \item In the second approach, we compute embeddings for each selected term in the document and compare these embeddings with the embeddings of neighboring terms within each sentence. To address variations in similarity values due to different contexts, we aggregate similarity values for each term pair across multiple occurrences in various contexts. By averaging these similarity values, we obtain a more stable measure of the semantic relationship between terms, which helps capture their general semantic similarity across diverse contexts.
\end{itemize}

In both cases, if the cosine similarity between two elements is higher than a predefined threshold value \textit{s}, nodes are created for these elements in the local proximity graph and an edge is created between them. The edge weight will be set to \textit{s} correspondingly. Thus, the hyperparameter \textit{s} regulates the growth of the term graph. This way, the used LLM acts as a reference resource.

When updating the graph with a new document, a nuanced approach is adopted to account for potential overlaps with existing term pairs. Each term pair from the new document is compared with those already in the graph. If a term pair overlaps, its overall similarity value is updated as follows: The edge's weight for that term pair is updated by averaging the existing similarity value(s), which need to be retained, with the new similarity. Here, it should be noted that the updated edge weight can be less than the threshold value \textit{s}. This approach can also be adapted so that only the maximum \textit{t} previously occurring term pair similarity values are included in this calculation in order to take into account the possibly important topicality and time dependency of the local documents. For non-overlapping term pairs, if their similarity surpasses the threshold \textit{s}, they are added to the graph as laid out before. This method balances retaining meaningful edges while accommodating new semantic associations, ensuring the graph's relevance and accuracy with continuously updated information. 

We also would like to point out that the partaking peers do not necessarily have to agree upon a common LLM. On the contrary, depending on the individual web servers' content, domain-specific models such as SciBERT \cite{Beltagy:19}, and FinBERT \cite{Araci:19} might be better suited to capture domain-specific characteristics and nuances. 

To incorporate new documents and cluster files into the \textit{WebMap} and establish links between documents and their corresponding cluster files, the following procedures in Algorithm~\ref{alg:linkind} must be executed:
\begin{algorithm}
\caption{Link induction and cluster assignment in the \textit{WebMap} }\label{alg:linkind}
\begin{algorithmic}[1]
    \STATE \textless START\textgreater
    \STATE Create local term proximity graph based on chosen LLM
    \STATE Cluster assignment: for all local documents, derive TRCs (cluster identification) based on local term proximity graph
    \FOR{all local documents} 
        \IF{respective cluster exists on the \textit{WebMap}}
            \STATE Attach document link to cluster
        \ELSE
            \STATE \label{itemx} Create cluster file locally and attach document link to cluster file
            \STATE  Derive the shortest path from the set of existing cluster files represented in the local term proximity graph to the new cluster file
            \STATE Create cluster files and bi-directional links among them for all nodes on the path
        \ENDIF
    \ENDFOR
        
    \STATE \textless END\textgreater
\end{algorithmic}
\end{algorithm}

The updated methodology, leveraging LLMs for inducing local term proximity graphs described before, is evident in steps 2 and 3.

\subsection{Towards a Semantic Signpost}

The assignment of individual documents to the global clusters is done by calculating their TRC terms using the local term proximity graphs just discussed. Within the global cluster files, it is now possible to organize the documents based on their similarity as well as in relation to incoming search queries, which are assigned TRCs, too. However, documents within a cluster will not only exhibit (flat, horizontal) semantic similarities, but topical (hierarchical) dependencies as well. For instance, a document on the main topic \textit{earthquake} could refer to contents that predominantly discuss its important subtopics such as \textit{seismic waves} and \textit{movement of plate boundaries}. 

The establishment of this intra-cluster linking structure yields a semantic signpost aiming to facilitate the targeted navigation to a topical direction of interest by lexically and semantically chaining documents. The mentioned topical dependencies can be uncovered by creating a cluster's directed and document-specific term associations graphs and applying for instance an extended variant \cite{Kubek:20} of the HITS algorithm \cite{Kleinberg:99} on them that takes into account the term association strengths $Assn$, too. For this purpose, the formulae for the update rules of the HITS algorithm must be extended. The authority value of a node x can then be determined using formula~\ref{eq:hitsassn1}:

\begin{equation}
\label{eq:hitsassn1}
a(x)= \sum_{v\rightarrow x} h(v) \cdot Assn(v \rightarrow x) 
\end{equation}

The hub value of a node x can be calculated using formula~\ref{eq:hitsassn2}:

\begin{equation}
\label{eq:hitsassn2}
h(x)= \sum_{x \rightarrow w} a(w) \cdot Assn(x \rightarrow w)
\end{equation}

\begin{figure}
    \centering
    \includegraphics[width=0.99\linewidth]{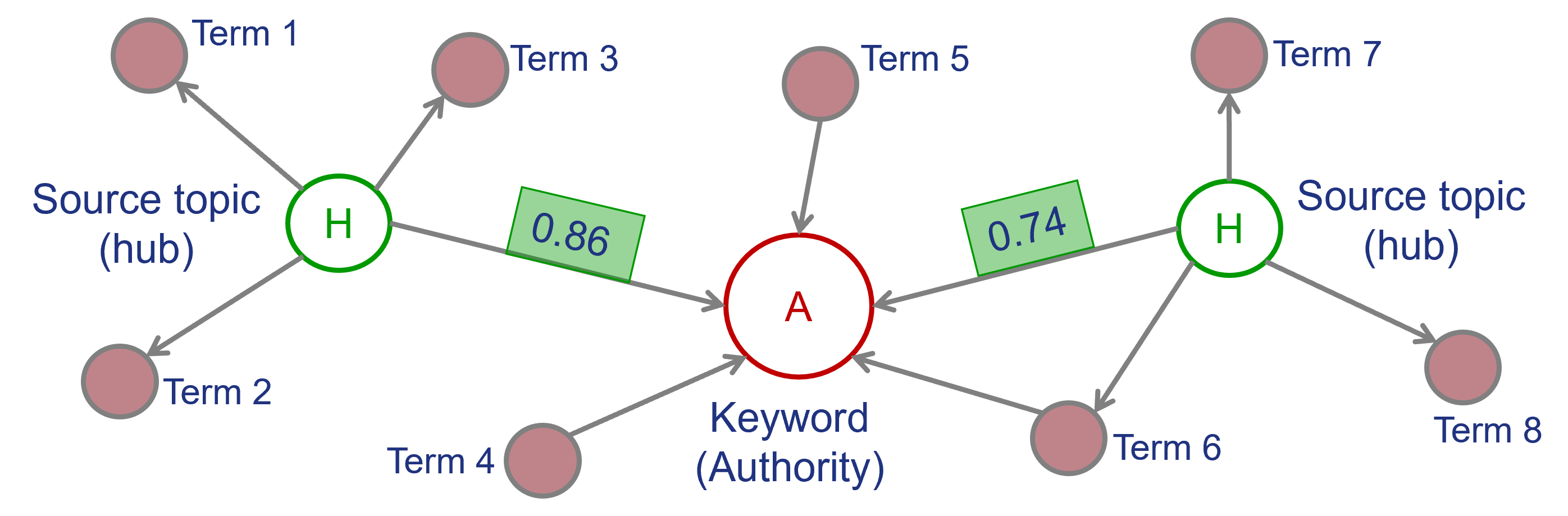}
    \caption{Determining authorities (keywords) and hubs (source topics) in directed term association graphs}
    \label{fig:Hits}
\end{figure}

This way, a document's main (authorities) and source topics (hubs) can be identified. Fig.~\ref{fig:Hits} illustrates this approach.

Table~\ref{tab:android} shows that the main topic ”Android” (mobile operating system) is greatly influenced by the subtopics ”source code”, ”development”, and ”platform”, which makes sense. 

\begin{table}
\caption{Terms with high authority and hub scores of the English Wikipedia article ”Android” (mobile operating system)}
\label{tab:android}
\centering
\begin{tabular}{lc}
\hline
\textbf{Term} & \textbf{Authority Score}\\
\hline
Android & 0.32 \\
Google & 0.31 \\
application & 0.27 \\ 
\hline
\end{tabular}
\begin{tabular}{lc}
\hline
\textbf{Term} & \textbf{Hub Score}\\
\hline
source code & 0.19 \\ 
development & 0.18 \\ 
platform & 0.14 \\ 
\hline
\end{tabular}
\end{table}

Based on the resulting lists of keywords and source topics per document, directed edges between the documents can now be induced as follows: 
If a document A primarily addresses the source topics of another document B, a link from A to B can be established. This exploits the fact that the source topics of documents are often the authorities of other documents, creating a chain of documents that are actually thematically dependent. This way, the herein described approach to obtain directed term associations is modified to gain the same effect at the document level, thereby generating recommendations for specific documents. In doing so, \textit{WebMap} incrementally learns new document relationships. Consequently, new search results can include links not only to similar documents but also to those primarily dealing with their source topics. This approach provides users access to background information on a topic of interest and allows them to follow related topics across multiple documents. The concept behind this approach is that the identified source topics, particularly, can guide users to documents covering important aspects of the analyzed search results, offering a novel method for discovering related documents.

\subsection{Detecting Subclusters and Handling Outliers}
The described cluster assignment process works in a sequential manner and the number of cluster files does not need to be specified beforehand. However, this number is constrained by the finite set of terms in a natural language.

Each cluster file is associated with a potentially large set of documents, which in turn can be associated with a set of subtopics and, in addition to the thematic dependencies described in the previous section, can provide important clues for vertical navigation within a global cluster file.
Therefore, it is advisable to regularly run an iterative and density-based clustering algorithm to identify those meaningful and disjoint subclusters. In the following considerations, we will simply refer to them as clusters.

\begin{figure}
	\centering
	\includegraphics[width=0.75\linewidth]{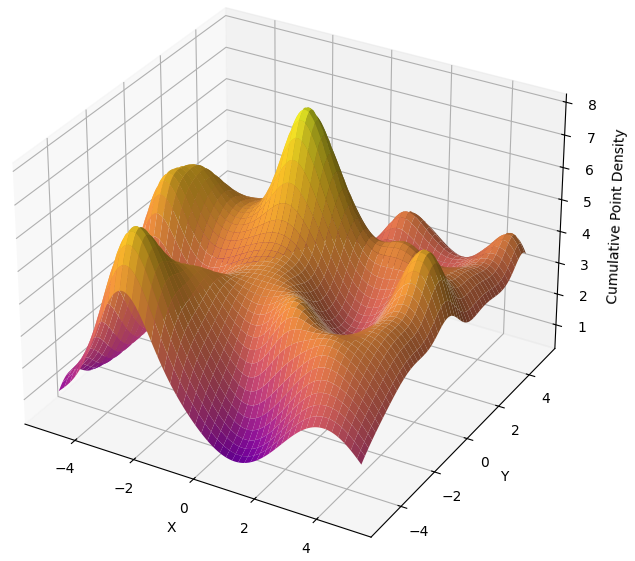}
	\caption{Multimodal distribution density (according to \cite{Bock:74})}\label{MmDen}
\end{figure}

In order to model the intrinsic characteristics of given sets of feature vectors with as low redundancy as possible, model-constituting clusters should be rather densely filled with data points -- our assigned documents -- and should have clear boundaries. An approach to do so is based on a probabilistic interpretation, which considers feature vectors of items (here: the documents) as observations of a mixed population constituted by several overlapping populations, the sum of whose single unimodal distribution densities is a multimodal distribution density, which has several local maxima (cf. Fig.~\ref{MmDen}). Under the condition, that the single populations are sufficiently separated, it is assumed, that the local maxima characterize the regions in feature space where the single populations are concentrated, i.e. where clusters are expected.

Based on this interpretation, the method proposed in \cite{Schnell:64} and its faster variant \cite{Komkhao:18} that replaces Gaussian functions by other bell-shaped and very similar looking curves, namely B-splines, are both able to detect clusters of complex shapes. According to the methods, those locations in feature space are searched, where a given data set exhibits local point concentrations with higher densities than in the respective vicinities. The search works by iteratively translating with a small step-size all feature vectors towards regions of higher point density. By this process, the vectors gradually approach the local maxima. Merging into a single cluster all feature vectors thus arriving in the neighbourhood of a certain location, an exhaustive and disjoint clustering of the data set is produced, with the number of these clusters derived from the characteristics of the data set, but not specified a priori.
This also facilitates the detection of outliers by identifying clusters with low point density after the algorithm's execution. 

In the context of \textit{WebMap}, documents in subclusters with a low point density can thus be regarded as outliers and as candidates for re-clustering. This could either mean that those documents will need to be assigned a different cluster file instead, or that they must be taken into account for a future subcluster assignment attempt. In any case, the described approach ensures that the subclusters obtained do not include any documents that are potentially semantically dissimilar within a cluster file. Likewise, a new management structure must be added to the cluster files that points to the subclusters and the documents contained in them.

\section{Discussion, Limitations, and Future Directions}
The proposed enhancements to \textit{WebMap} introduce significant improvements to the existing framework, aiming to address key challenges in web-based research activities. By leveraging LLMs, \textit{WebMap} seeks to enhance the semantic understanding of web content, thereby improving the accuracy and relevance of search results. The transition from co-occurrence graphs to local term proximity graphs induced by LLMs represents a paradigm shift in how document clusters are determined, enabling a more nuanced analysis of semantic relationships between terms and documents.

However, while the adoption of LLMs holds great promise for enhancing the effectiveness of \textit{WebMap}, several limitations and challenges need to be addressed. Firstly, the reliance on textual content restricts the applicability of \textit{WebMap} to domains where multimedia content plays a significant role. Future iterations of the system may need to explore methods for integrating and analyzing diverse media types to ensure comprehensive coverage of web-based information.

Moreover, the current implementation of intra-cluster signposts, while facilitating navigation within clusters, may hinder the exploration of cross-cluster relationships and thematic connections. Enhancements to the linking structure, possibly through the incorporation of semantic similarity measures or thematic analysis algorithms, could improve the discoverability of related content across different clusters.

To maximize the effectiveness of utilizing \textit{WebMap}, it is essential to dedicate effort towards analyzing ideal use cases. This involves examining how current user workflows, such as enhancing search capabilities for specific topics, implementing targeted navigation features, and optimizing content crawling and discovery processes to focus on relevant information, can be enhanced further.

In the context of developing innovative use cases through the utilization of \textit{WebMap's} technology, the newly devised approaches for sustainable research support can play a pivotal role. For instance, the herein discussed method of identifying sources discussing a given topic X, which exhibit high relevance, either hierarchically as subordinates or superordinates, or present contradictory perspectives, makes thematic dependencies explicit. \textit{WebMap} may also serve as a foundational search index for the development of novel search agents, especially those Artificial Intelligence (AI) tools aimed at the research synthesis process of identifying, organizing, extracting, and combining studies.
This is particularly significant as researchers often struggle to keep up with new findings, especially within their specific fields. This challenge is exacerbated by disciplinary silos, especially in interdisciplinary areas like STEM education research. For example, cognitive psychologists may miss relevant work in math education research, and vice versa. Therefore, AI tools need to be developed to help accelerate, transform and reimagine the practice of research synthesis while being user-friendly, reliable, trustworthy, ethical and possibly freely available.
Such tools can be suitably backed by \textit{WebMap's} search infrastructure.

Additionally, the reliance on cooperative peers for providing cluster files introduces challenges related to data redundancy, reliability, and distribution. Ensuring equitable access to cluster files and maintaining data integrity across distributed networks will be crucial for the scalability and robustness of \textit{WebMap} in real-world settings. To cope with the sheer mass of online text resources, it therefore seems to make the most sense to apply \textit{WebMap} and its herein discussed extensions to closed networks such as enterprise networks and intranets first.

Overall, while the proposed enhancements represent a significant step forward in enhancing web-based research capabilities, ongoing research and development efforts are needed to address the aforementioned limitations and ensure the effectiveness and usability of \textit{WebMap} across diverse use cases and environments. Collaborative efforts from the research community will be essential in advancing the state of the art in web-based information retrieval and knowledge discovery.

\section{Conclusion}
The proposed extensions to \textit{WebMap} aim to support web-based research activities by leveraging advanced natural language processing techniques. By integrating LLMs and refining clustering algorithms, \textit{WebMap} aims to provide users with more accurate, relevant, and comprehensive search results, ultimately enhancing their ability to navigate and explore complex information spaces on the web.

However, to fully realize the potential of \textit{WebMap}, it is imperative to address the existing limitations and challenges, including the restriction to textual content, the need for improved cross-cluster navigation, and the complexities associated with distributed data management. Through continued research and innovation, \textit{WebMap} has the potential to become a valuable tool for researchers, educators, and information seekers, facilitating seamless access to knowledge and insights across the vast expanse of the World Wide Web. These aspects will be addressed in future research works.

%
%
%

\begin{thebibliography}{8}
\bibitem{Rossrucker:24}
Roßrucker, G.: A Concept for a Distributed WebMap. Springer Cham (2024)

\bibitem{Kubek:16}
Kubek, M., Unger, H.: Centroid terms as text representatives. In: Proceedings of the 2016 ACM Symposium on Document Engineering, pp. 99--102, ACM, New York, NY, USA (2016)

\bibitem{Jin:07}
Jin W., Srihari R. K.: Graph-based text representation and knowledge discovery. In: Proceedings of the 2007 ACM Symposium on Applied Computing, ACM, New York, NY, USA (2007)

\bibitem{Biemann:22}
Biemann, C., Heyer, G., and Quasthoff, U.: Wissensrohstoff Text: Eine Einf{\"u}hrung in das Text Mining. 2nd Edition, Springer Fachmedien Wiesbaden (2022)

\bibitem{Devlin:19}
Devlin, J., Chang, M., Lee, K., and Toutanova, K.: Bert: Pre-training of deep bidirectional transformers for language understanding. (2019)

\bibitem{Beltagy:19}
Beltagy, I., Lo, K., and Cohan, A.: Scibert: A pretrained language model for scientific text. (2019)

\bibitem{Araci:19}
Araci, D.: Finbert: Financial sentiment analysis with pre-trained language models. (2019)

\bibitem{Kubek:20}
Kubek, M.: Concepts and Methods for a Librarian of the Web. In: Studies in Big Data, Volume 62, Springer, Cham (2020)

\bibitem{Kleinberg:99}
Kleinberg, J. M.: Authoritative sources in a hyperlinked environment. In J. ACM, volume 46, pp. 604--632, ACM, New York, NY, USA (1999)

\bibitem{Bock:74} 
Bock, H. H.: Automatische Klassifkation. G\"ottingen: Vandenhoeck \& Ruprecht (1974)

\bibitem{Schnell:64}
Schnell, P.: Eine Methode zur Auffindung von Gruppen. In: Biometrische Zeitschrift, 6, pp. 47--48 (1964)

\bibitem{Komkhao:18} 
Komkhao, M., Kubek, M., and Halang, W. A.: Sequential clustering and condensing the meaning of texts into centroid terms. In Information Technology Journal, volume 14, pp. 1--10 (2018)

\bibitem{Vaswani:17} 
Vaswani, A. et al.: Attention Is All You Need. arXiv:1706.03762v7 [cs.CL] (2017)

\bibitem{Minaee:24}
Minaee, S., Mikolov, T. et. al.: Large Language Models: A Survey. arXiv:2402.06196v2 [cs.CL] (2024)

\bibitem{Brown:20}
Brown, T. et al., Language models are few-shot learners. Advances in neural information processing
systems, vol. 33, pp. 1877--1901 (2020)

\bibitem{GPT4:23}
OpenAI: GPT-4 Technical Report. https://arxiv.org/pdf/2303.08774v3.pdf (2023)

\bibitem{laama1:23}
Touvron, H. et al.: Llama: Open and efficient foundation language models. arXiv:2302.13971 [cs.CL] (2023)

\bibitem{laama2:23}
Touvron, H., Martin, L. et al.: Llama2: Open foundation and fine-tuned chat models. arXiv:2307.09288 [cs.CL] (2023)

\bibitem{Jiang:23}
Jiang, A. Q., Sablayrolles, A. et.al.: Mistral 7B.  arXiv:2310.06825v1 [cs.CL] (2023)


\end{thebibliography}
%

\end{document}